\documentclass[%
 aps,
 jmp,%
 amsmath,amssymb,
 reprint,%
]{revtex4-1}

\usepackage[pdftex]{graphicx}
\usepackage{dcolumn}
\usepackage{bm}
\usepackage[hidelinks]{hyperref}

\begin{document}

\preprint{APS/123-QED}

\title{Cold Brownian motion in aqueous media via anti-Stokes photoluminescence}

\author{Paden B. Roder}\thanks{These authors contributed equally to this work.}
\affiliation{Department of Materials Science and Engineering, University of Washington, Seattle USA 98195}%
\author{Bennett E. Smith}\thanks{These authors contributed equally to this work.}
\affiliation{Department of Chemistry, University of Washington, Seattle, WA 98195}%
\author{Xuezhe Zhou}\thanks{These authors contributed equally to this work.}
\affiliation{Department of Materials Science and Engineering, University of Washington, Seattle USA 98195}%
\author{Matthew J. Crane}
\affiliation{Department of Chemical Engineering, University of Washington, Seattle, WA 98195}%
\author{Peter J. Pauzauskie}\thanks{Corresponding author: peterpz@uw.edu}
\affiliation{Department of Materials Science and Engineering, University of Washington, Seattle USA 98195}%
\affiliation{Fundamental Computational Sciences Directorate, Pacific Northwest National Laboratory, Richland, WA 99354}%

\date{\today}

\begin{abstract}
Advances in cryogenic sciences have enabled several observations of new low-temperature physical phenomena including superconductivity, superfluidity, and Bose-Einstein condensates. Heat transfer is also critical in numerous applications including thermal management within integrated microelectronics and the regulation of plant-growth and development. Here we demonstrate that single-beam laser-trapping can be used to induce and quantify the local refrigeration of aqueous media through analysis of the cold Brownian dynamics of individual Yb$^{3+}$-doped yttrium lithium fluoride (YLF) crystals in an inhomogeneous temperature field via forward light scattering and back-focal-plane interferometry. A tunable, NIR continuous-wave laser is used to optically trap individual YLF crystals with an irradiance on the order of 1 MW/cm$^2$. Heat is transported out of the crystal lattice (across the solid / liquid interface) by anti-Stokes photoluminescence following upconversion of Yb$^{3+}$ excited states mediated by optical-phonon absorption. The cold Brownian motion (CBM) analysis of individual YLF crystals indicates local cooling by $>$21$^\circ$C below ambient conditions suggesting a range of potential future applications.
\end{abstract}

\maketitle

\section{\label{sec:Intro}Introduction}
In 1929, Pringsheim\cite{pringsheim1929-8} proposed that solid-state materials could experience refrigeration if they exhibited biased emission of anti-Stokes (blue-shifted) radiation relative to a fixed optical excitation wavelength. Epstein\cite{epstein1995observation-9} and colleagues experimentally demonstrated this concept first in 1995 using rare-earth-doped fluoride glass materials (ZBLAN). More recently, it has been shown that solid-state rare-earth-doped yttrium-lithium-fluoride (Yb$^{3+}$:YLF) crystals can be cooled to cryogenic temperatures\cite{melgaard2013optical-10} \textit{in vacuo} using a continuous-wave NIR laser excitation. Furthermore, the laser refrigeration of doped yttrium-aluminum-garnet (Yb$^{3+}$:YAG) materials has recently been reported in air at atmospheric pressure\cite{nemova2013laser-11}. Anti-Stokes photoluminescence has also been shown\cite{zhang2013laser-12} to cool cadmium sulfide (CdS) nanoribbons \textit{in vacuo} by as much as 40 $^\circ$C below room temperature.

In contrast to anti-Stokes processes, optomechanical laser refrigeration has also been demonstrated based on a novel mechanism of angular momentum transfer between a circularly polarized laser and a birefringent crystal\cite{arita2013laser-13}. To date, laser-refrigeration of nanocrystals in aqueous media has not been reported stemming primarily from the large near-infrared (NIR) optical absorption coefficient of water\cite{palik1998handbook-14}. It has remained an open question whether these known cooling materials could act to refrigerate aqueous media and undergo hypothesized cold Brownian motion (CBM)\cite{chakraborty2011generalised-7,kroy2014levitating-15}. In this work, we demonstrate the local refrigeration of aqueous media by $>$ 21 $^\circ$C below ambient conditions following anti-Stokes photoluminescence from optically trapped\cite{ashkin1990force-6}, rare-earth-doped YLF nanocrystals undergoing CBM.
\section{\label{sec:resultsdiscuss}Results \& Discussion}
\subsection{\label{sec:synchar}Nanocrystal Synthesis \& Characterization}
\begin{figure*}[t]
	\begin{center}
		\includegraphics[width=12cm]{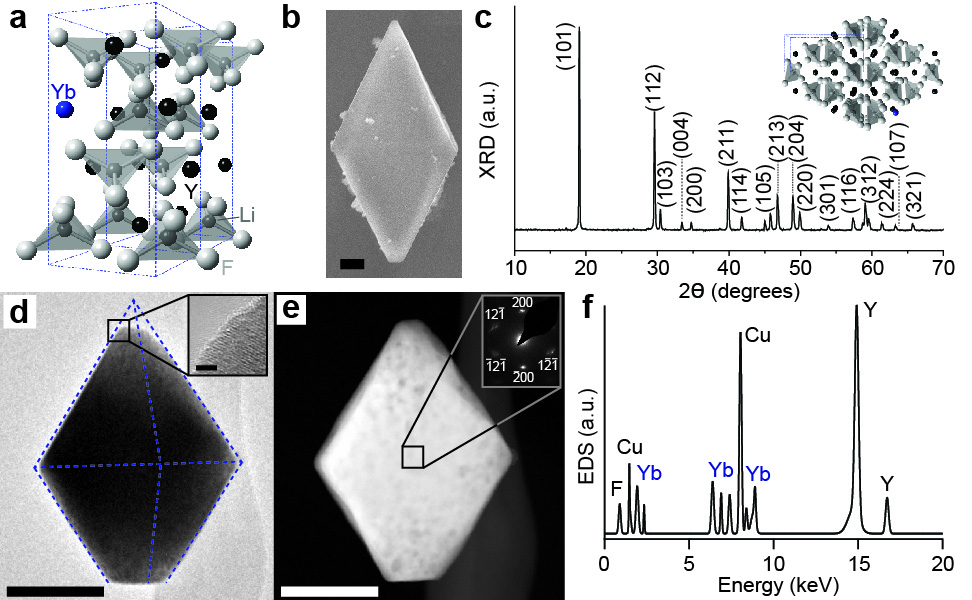}
	\end{center}
	\caption{\textbf{Synthesis and characterization of YLF crystals. a,} Schematic of Scheelite crystal structure of YLF with I$_{41/a}$ space group symmetry. \textbf{b,} Scanning electron microscope image of a faceted (Yb$^{3+}$)$_{0.1}$(Y$^{3+}$)$_{0.9}$LiF$_4$ particle exhibiting TTB morphology. Scale bar = 1 $\mu$m. \textbf{c,} Powder x-ray diffraction pattern of YLF crystals following hydrothermal synthesis indicating a pure Scheelite crystal phase. Inset: schematic of TTB morphology relative to YLF's unit cell. \textbf{d,} Bright field transmission electron microscope (TEM) image of an individual Yb$^{3+}$:YLF grain; scale bar = 200 nm. Inset: high-resolution TEM image taken from the indicated region; scale bar = 2 nm. \textbf{e,} High-angle annular-dark-field (HAADF) image of the YLF grain in panel b showing regions of high contrast suggesting the presence of polycrystalline domains. Inset: select area electron diffraction from the indicated region. \textbf{f,} X-ray fluorescence compositional-analysis-spectrum of an individual YLF crystal taken within the TEM confirming the elemental crystalline composition including Y, Yb, and F species.}
	\label{fig:character}
\end{figure*}
Pioneering efforts to cool Yb$^{3+}$:YLF materials \textit{in vacuo} have relied on the growth of high-purity YLF single-crystals using an air- and moisture- free Czochralski process\cite{seletskiy2010laser-16}. In the experiments reported here, a low-cost modified hydrothermal synthesis\cite{lu2011hydrothermal-17} of Yb$^{3+}$:YLF is used to prepare YLF crystals shown in Fig. \ref{fig:character}. Scanning electron microscopy reveals that YLF crystals exhibit a truncated tetragonal bipyramidal (TTB) morphology (Fig. \ref{fig:character}b). X-ray diffraction shows that the YLF crystal has a Scheelite structure (Fig. \ref{fig:character}c). Bright field / HAADF TEM imaging (Fig. \ref{fig:character}d/e) and electron diffraction suggest that the TTB materials are polycrystalline and likely form through an oriented attachment\cite{li2012direction-18} process of nanocrystalline grains (Fig. \ref{fig:character}e, inset).

\subsection{Optical Trap Setup}
\begin{figure*}[t]
	\begin{center}
		\includegraphics[width=10cm]{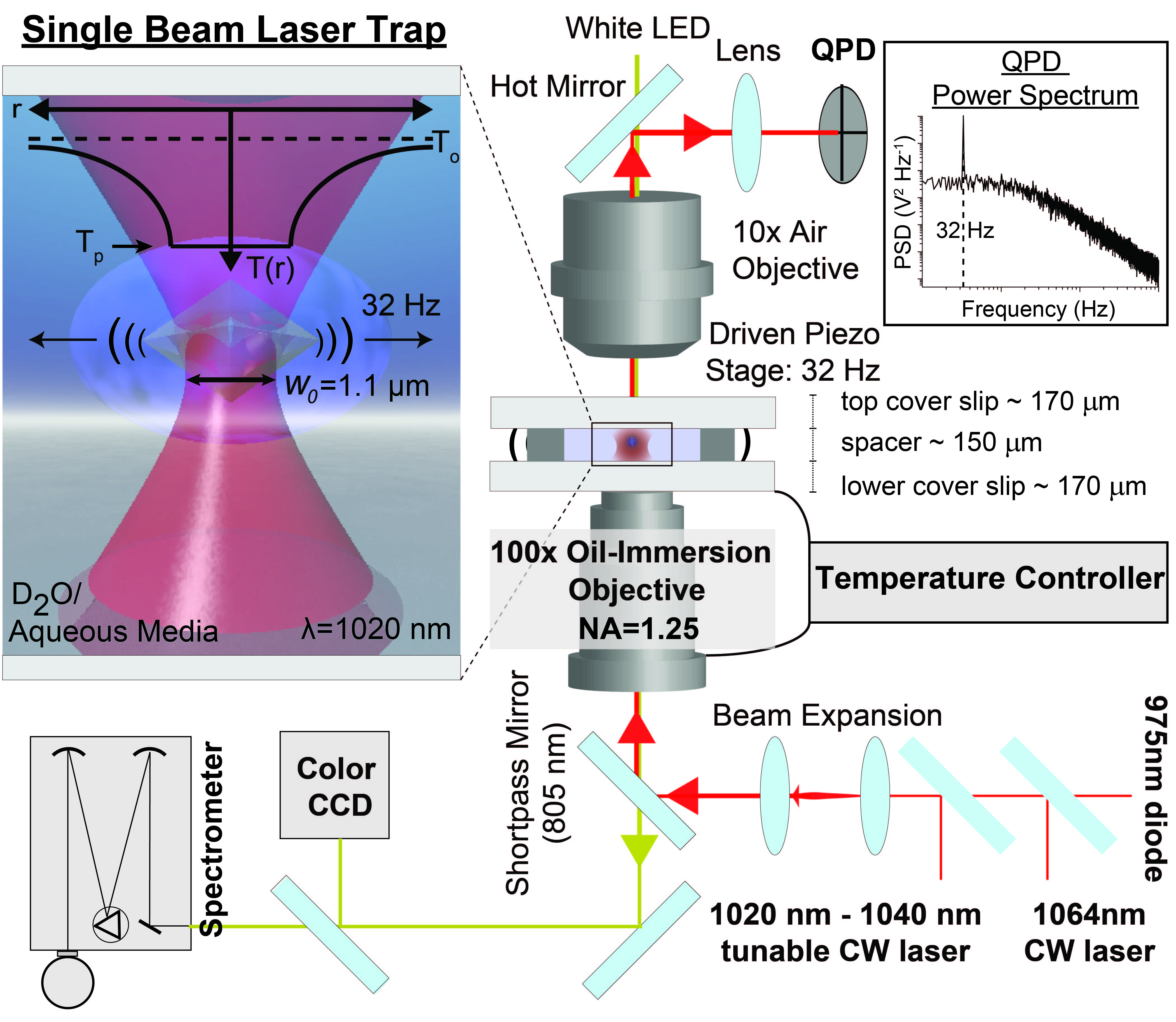}
	\end{center}
	\caption{\textbf{Schematic of laser trapping instrument.} An optically trapped YLF crystal in an aqueous fluid chamber. A piezostage driven at 32 Hz produces a peak in the quadrant photodiode (QPD) power spectrum which is used to extract a calibrated diffusion constant. The particle's temperature (T$_p$) and local temperature profile is then extracted using cold Brownian motion analysis.}
	\label{fig:system}
\end{figure*}

A laser trapping system shown in Fig. \ref{fig:system} was used to optically trap individual Yb$^{3+}$:YLF crystals. Briefly, a single-beam laser trap was used to extract the surrounding local temperature profile of optically trapped rare-earth-doped YLF particles by analysis of calibrated power spectra of the forward-scattered laser radiation\cite{roder2014photothermal-19}, which is used to quantify the CBM of the particle. In order to minimize fluid absorption and heating at the trapping wavelengths\cite{haro2013optical-20}, laser trapping experiments were performed in D$_2$O unless explicitly stated otherwise. The laser trap setup is outlined in Materials \& Methods, and the CBM temperature analysis is described in Appendix \ref{sec:AppI}.

\subsection{CBM of Locally Refrigerated Nanocrystals}
\begin{figure}[h]
	\begin{center}
		\includegraphics[width=8cm]{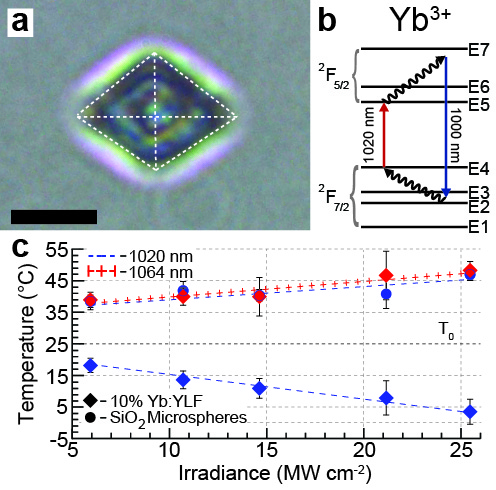}
	\end{center}
	\caption{\textbf{Laser refrigeration of optically trapped YLF microcrystals. a,} Optical micrograph of an optically trapped YLF crystal; scale bar = 3 $\mu$m. \textbf{b,} Crystal field energy level configuration of Yb$^{3+}$ dopant ions and employed cooling scheme. \textbf{c,} Extracted temperature of optically trapped particles in D$_2$O. Yb$^{3+}$-doped YLF particles are shown to cool when trapping wavelength is resonant with the E4-E5 transition ($\lambda_{trap}$ = 1020 nm) but heat when the trapping wavelength is below the transition ($\lambda_{trap}$ = 1064 nm). As a control, amorphous silica microspheres heat when trapped at $\lambda_{trap}$ = 1020 nm.}
	\label{fig:resultsCBM}
\end{figure}

A bright-field micrograph for a characteristic optically-trapped Yb$^{3+}$:YLF crystal is shown in Fig. \ref{fig:resultsCBM}a. The dependence of laser refrigeration on the trapping laser’s pump wavelength is shown in Fig. \ref{fig:resultsCBM}c, where YLF crystals doped with 10\% Yb$^{3+}$ are observed to cool from 19 $^\circ$C at a 5.9 MW/cm$^2$ trapping irradiance to 4 $^\circ$C at a 25.5 MW/cm$^2$ trapping irradiance when trapped with $\lambda$ = 1020 nm, which is resonant with ytterbium's E4-E5 transition shown in Fig. \ref{fig:resultsCBM}b. The same Yb$^{3+}$:YLF crystals are shown to heat from 40 $^\circ$C to 47 $^\circ$C when trapped at the same respective irradiances with $\lambda$ = 1064 nm, which is energetically insufficient to pump the E4-E5 resonance and subsequently cannot initiate upconversion-mediated cooling. As a control, 1.01 $\mu$m diameter amorphous silica microspheres are observed to heat from 39 $^\circ$C to 46 $^\circ$C when trapped with the $\lambda$ = 1020 nm trapping wavelength.

\subsection{Ratiometric Thermometry of Codoped Nanocrystals}
Furthermore, it has been recently reported that strong visible upconversion in rare-earth codoped nanocrystals can be used for efficient biological imaging and labeling\cite{wang2010upconversion-21}. The long (ms) lifetime of photoluminescence from rare-earth point defects may also be used to infer temperature changes through ratiometric thermometry by analysis of Boltzmann thermal populations\cite{mclaurin2013dual-22}, which is given by the equation:
\begin{equation}
\frac{I_2}{I_1}\propto\exp\left(\frac{-(E_2 - E_1)}{k_bT}\right)
\label{eq1}
\end{equation}

\begin{figure*}[t]
	\begin{center}
		\includegraphics[width=12cm]{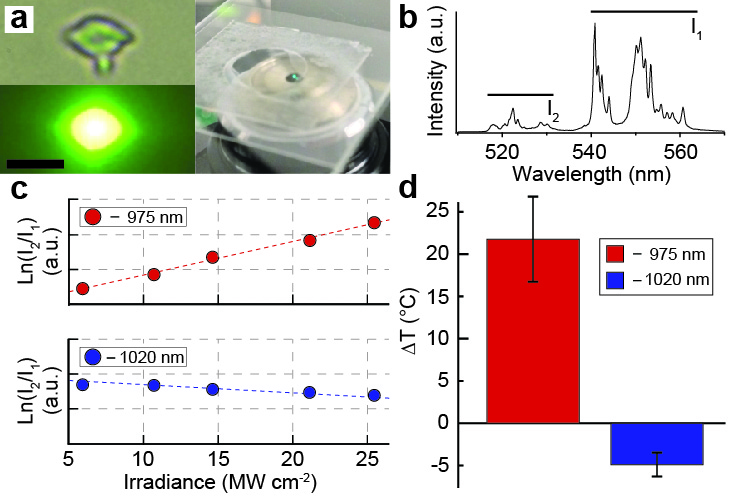}
	\end{center}
	\caption{\textbf{Upconversion and ratiometric thermometry of codoped YLF. a,} Bright-field optical micrograph showing a codoped 2\%Er$^{3+}$,10\%Yb$^{3+}$:YLF particle in Brownian motion (top-left) and a dark-field optical micrograph of the crystal when trapped with $\lambda_{trap}$ = 1020 nm (bottom-left). Scalebar = 4$\mu$m. Upconverted photoluminescence can be seen with the unaided eye (right). \textbf{b,} Photoluminescence spectra of the corresponding dark-field image showing the integration regions I$_2$ and I$_1$, representing emission from Er$^{3+}$ energy states E$_2$ ($^2$H$_{11/2}$) and E$_1$ ($^4$S$_{3/2}$) to the ground state E$_{ground}$ ($^4$I$_{15/2}$), respectively. \textbf{c,} Natural logarithm of the ratio I$_2$/I$_1$ showing a linear increase (top) with laser irradiance at $\lambda_{trap}$ = 975 nm and a linear decrease (bottom) with laser irradiance at $\lambda_{trap}$ = 1020 nm. \textbf{d,} Laser refrigeration of the codoped YLF crystal analyzed in c measured via cold Brownian motion analysis.}
	\label{fig:resultsRatio}
\end{figure*}

\noindent In brief, changes in the ratio of the integrated emission bands $I_2$ and $I_1$ that stem from transitions between energy states $E_2$ and $E_1$, respectively, and a common ground state are directly correlated to a change in the particle's temperature. Codoping YLF crystals with both Yb$^{3+}$ and Er$^{3+}$ ions leads to a thermalized Boltzmann distribution between the E$_2$ ($^2$H$_{11/2}$) and E$_1$ ($^4$S$_{3/2}$) manifolds of Er$^{3+}$ and an intense green upconversion emission that is visible to the unaided eye, as shown in Fig. \ref{fig:resultsRatio}a.

Utilizing photoluminescence spectroscopy with optically trapped codoped upconverting nanocrystals provides a unique capability of testing the particle-to-particle variability in emission. For the codoped 2\%Er$^{3+}$, 10\%Yb$^{3+}$:YLF particles reported, substantial fluctuations in upconversion photoluminescence was observed, making bulk calibration measurements inadequate for individual ratiometric temperature measurements as detailed in Appendix \ref{sec:AppIII}. Thus, ratiometric thermometry can be used only qualitatively during laser trapping experiments to observe changes in temperature as the trapping irradiance is increased, as shown in Fig. \ref{fig:resultsRatio}b/c. The increase (decrease) in the logarithmic ratio of I$_2$ to I$_1$ (Fig. \ref{fig:resultsRatio}b/c) with increasing irradiance reflects an increase (decrease) in the internal lattice temperature\cite{vetrone2010temperature-23}, which agrees well with the observed temperature changes measured via laser trapping light scattering temperature analysis (Fig. \ref{fig:resultsRatio}d).
\noindent Specifically, laser trapping analysis of the particles' CBM indicates that codoped 2\%Er$^{3+}$, 10\%Yb$^{3+}$:YLF undergoes laser refrigeration ($\Delta$T = -4.9 $\pm$ 2.8 $^\circ$C) when trapped at $\lambda$ = 1020 nm and heating ($\Delta$T = 21.8 $\pm$ 10.11 $^\circ$C) when trapped at $\lambda$ = 975 nm. Furthermore, it has been proposed recently that codoping YLF crystals with other upconverting rare-earth ions can enhance cooling through energy transfer enhanced cooling\cite{dong2013energy-24}. The laser trapping and refrigeration of 10\%Yb$^{3+}$:YLF nanocrystals at $\lambda$ = 1020 nm in biological buffers such as phosphate-buffered saline (PBS) and Dulbecco's Modified Eagle Medium (DMEM) is also possible, as demonstrated in Table \ref{tab:solvents}.%
\begin{table}[h]
\caption{\textbf{Local cooling of Yb$^{3+}$-doped YLF crystals in various media.}}
\label{tab:solvents}
\begin{tabular}{|c||c|c|c|}
\toprule
\textbf{Solvent} & \textbf{$\Delta$T = (T$_p$ - T$_0$) ($^\circ$C)} & \textbf{$\Delta$T Std. Dev. ($^\circ$C)}\\
\hline
D$_2$O & -15.0 & 4.1 \\
DI Water & -14.7 & 3.8 \\
PBS & -14.9 & 4.3 \\
DMEM & -11.2 & 6.3 \\
\hline%
\end{tabular}%
\def\tablename{}%
\def\thetable{}%
\caption{\footnotesize{$^*$ Temperature changes are determined between the laser irradiances 5.9 and 25.5 MW/cm$^2$, where T$_0$ = 25 $^\circ$C.}}
\end{table}
\subsection{Outlook}
These results illustrate the potential of using singly- and co-doped YLF nanocrystals as a platform for physiological refrigeration, biological imaging, and \textit{in situ} ratiometric thermometry. Potential applications for these materials include triggering and probing fundamental temperature-controlled biological processes\cite{lee2013regulation-5,kucsko2013nanometre-25}. The ability to measure and to control temperature could enable this technology to investigate kinetics and temperature sensitivity of basic cellular processes, including ion channels\cite{huang2010remote-26}, or to act as simultaneous theranostic-hypothermia agents to identify and treat cancerous tissues\cite{li2014targeted-27}.

Analyzing the CBM of a nanocrystal dispersed in a liquid phase to measure the nanocrystal's temperature also provides the unique capability to extract the local temperature gradient in the medium surrounding the trapped nanocrystal. Since the aspect ratio of the TTB materials used here is near unity, we approximate the radius R of the particles using an equivalent-sphere model and can extract the local temperature field a distance r from the particles' surface (at temperature T$_p$, excluding the temperature discontinuity at the particle's surface from the Kapitza resistance\cite{cahill2003nanoscale-4}), which is given by\cite{rings2010hot}:
\begin{equation}
T(r) = T_0 + \frac{R}{r}(T_p - T_0)
\label{eq2}
\end{equation}

\noindent where $T_0$ is the bath temperature of the medium. Given that the average radius of the Yb$^{3+}$:YLF particles trapped at $\lambda$ = 1020 nm in Fig. \ref{fig:resultsCBM}c is R$_{avg}$ = 764 $\pm$ 293 nm, T$_0$ = 25 $^\circ$C, and T$_{p,avg}$ = 3.4 $^\circ$C at 25.5 MW/cm$^2$ irradiance, the distance away from the particle where the temperature increases to within 1\% of T$_0$ is 6.9 $\mu$m (Fig. \ref{fig:system}). Future synthetic developments with YLF host crystals could enhance the resonant optical absorption of Yb$^{3+}$ through morphology dependent resonances\cite{roder2012nanowire-28}, or reduce unwanted background impurity absorption to a level capable of locally freezing water at atmospheric or elevated pressures.%
\section{\label{sec:matmeth}Materials \& Methods}%
\subsection{YLF Synthesis}%
The following synthesis was performed following modifications to Lu, C. et. al\cite{lu2011hydrothermal-17}. Yttrium oxide (Y$_2$O$_3$), ytterbium oxide (Yb$_2$O$_3$) and erbium oxide (Er$_2$O$_3$) are of 99.99\% purity and used as purchased from Sigma-Aldrich. Yttrium nitrate (Y(NO$_3$)$_3$), ytterbium nitrate (Yb(NO$_3$)$_3$) and erbium nitrate (Er(NO$_3$)$_3$) are obtained by dissolving the oxide in concentrated nitric acid at 60 $^\circ$C while stirring for several hours until excess nitric acid is removed. The residual solid is then dissolved in Millipore deionized (DI) water to achieve a stock concentration of the respective nitrate. Lithium fluoride (LiF), nitric acid (HNO$_3$), ammonium bifluoride (NH$_4$HF$_2$) and ethylenediaminetetraacetic acid (EDTA) are analytical grade and used directly in the synthesis without any purification. The following preparation uses the synthesis of 2\%Er$^{3+}$,10\%Yb$^{3+}$:YLiF$_4$ as an example. 7.04ml of 0.5M Y(NO$_3$)$_3$, 0.8ml of 0.5M Yb(NO$_3$)$_3$ and 0.16ml of 0.5M Er(NO$_3$)$_3$ are mixed with 1.17g EDTA in 5ml Millipore DI water at 80 $^\circ$C while stirring for 1h. This is solution A. Subsequently, 0.21g of LiF and 0.68g of NH$_4$HF$_2$ are dissolved in 7ml Millipore DI water at 70 $^\circ$C while stirring for 1h to form solution B. Solutions A and B are mixed together while stirring for 20min to form a homogeneous white suspension which is then transferred to a 23ml Teflon-lined autoclave and heated to 220 $^\circ$C for 72h. After the autoclave cools to room temperature, the 2\%Er$^{3+}$,10\%Yb$^{3+}$:YLiF$_4$ particles can be recovered by centrifuging and washing with ethanol and Millipore DI water three times. The final white powder is obtained by calcining at 300 $^\circ$C for 2h. 10\%Yb$^{3+}$:YLiF$_4$ particles are achieved using the same method.%
\subsection{TEM Characterization}%
Bright field and STEM HAADF images were taken on a FEI Tecnai G2 F20 at an accelerating voltage of 200 keV. Select area electron diffraction (SAED) images were taken with a camera length of 490 mm. EDS spectra were obtained with a 60 second acquisition time. The spectra were then processed by subtracting the background and smoothing the peaks.%
\subsection{SEM Characterization}%
Secondary electron images were taken on an FEI Sirion at an accelerating voltage of 5 keV.%
\subsection{XRD Characterization}%
Powder x-ray diffraction (XRD) patterns are obtained on a Bruker F8 Focus Powder XRD with Cu Kɑ (40kV, 40mA) irradiation ($\lambda$=0.154nm). The 2$\theta$ angle of the XRD spectra is from 10$^\circ$ to 70$^\circ$ and the scanning rate is 0.01$^\circ$s$^{-1}$. The one minor unlabeled peak in the XRD spectra at 2$\theta$ = 44.9$^\circ$ is attributed to a small amount of unreacted LiF precursor ((200) peak).%
\subsection{Laser Trapping Description}
The laser tweezer setup is a modified modular optical tweezer kit (Thorlabs, OTKB), where the original condenser lens has been replaced with a 10x Mitutoyo condenser (Plan Apo infinity-corrected long WD objective, Stock No. 46-144). The 100x objective focusing lens has a numerical aperture of 1.25 and a focal spot of 1.1 $\mu$m. The quadrant photodiode and piezostage were interfaced to the computer through a DAQ card (PCIe-6361 X Series, National Instruments) and controlled through modified MATLAB software (Thorlabs). Experimental chambers were prepared as follows. Several microliters of the nanocrystal/aqueous medium dispersion were transferred by a pipette into a chamber consisting of a glass slide and glass coverslip. The edges of the glass slide and the glass coverslip were then sealed with a 150-$\mu$m-thick adhesive spacer (SecureSeal Imaging Spacer, Grace Bio-labs).  Nanocrystals were trapped at the center (~75 $\mu$m from the surface) of the temperature controlled perfusion chamber (RC-31, Warner Instruments) and held at T$_0$ = 25 $^\circ$C while voltage traces were recorded at the quadrant photodiode (QPD) for 3 seconds at a sample rate of 100 kHz. The QPD voltage signal was calibrated by oscillating the piezostage at 32Hz and an amplitude of 150nm peak-to-peak during signal acquisition, as outlined in reference\cite{tolic2006calibration-29}. Trapping data was acquired using a diode-pumped solid state Yb:YAG thin-disk tunable laser (VersaDisk 1030-10, Sahajanand Laser Technologies) at a wavelength of 1020 nm, a 975 nm pigtailed Fiber Bragg Grating (FBG) stabilized single-mode laser diode (PL980P330J, Thorlabs), as well as a solid-state Nd:YAG 1064 nm (BL-106C, Spectra-Physics) at an irradiance of 5.9, 10.7, 14.6, 21.2,  and 25.5 MW/cm$^2$. Each YLF cooling data point in Fig. \ref{fig:resultsCBM}c in the manuscript represents an average of 6 individual particles with an average radius of 764 nm with a standard deviation of 293 nm. Silica beads (SS04N/9857, Bangs Laboratories) were used for their monodisperse size distribution (1010 nm diameter), and they have shown to minimally heat when trapped with a laser tweezer at NIR wavelengths\cite{peterman2003laser-30}. Electromagnetic simulations of the interaction of the trapping laser with a YLF TTB were also performed to predict the stable trapping configurations of optically trapped YLF particles, detailed in Appendix \ref{sec:AppV}. Lastly, visible emission of Er$^{3+}$ from Er/Yb codoped trapped YLF host crystals was detected using an Acton SpectraPro 500i spectrograph with a Princeton liquid-nitrogen cooled Si detector.%

\section{\label{sec:Acknowledge}Acknowledgements}
This research was made possible by a grant from the Air Force Office of Scientific Research Young Investigator Program (contract \#FA95501210400), start-up funding from the University of Washington, as well as a capital equipment donation from the Lawrence Livermore National Laboratory. P.B.R. thanks the NSF for a Graduate Research Fellowship under grant number DGE-1256082. M.J.C was supported by the Department of Defense (DoD) through the National Defense Science \& Engineering Graduate Fellowship (NDSEG) Program. The authors thank Klaus Kroy of Leipzig University for discussion of CBM analysis, John W. Cahn for discussion of YLF crystallography, and E. James Davis for manuscript comments and providing an optical spectrometer with LN$_2$-cooled detector.%

\section{\label{sec:AuthorContrib}Author Contributions}
X.Z. performed the hydrothermal synthesis and characterization (XRD, SEM, photoluminescence) of YLF materials as well as performed laser trapping experiments. B.E.S. contributed to operation of the tunable solid-state trapping laser, visible spectroscopy, laser trapping experiments, as well as computational electromagnetics calculations using DDSCAT. P.B.R. performed TEM analysis of YLF materials, helped with operations and alignment of the tunable trapping laser, conducted laser trapping experiments, as well as wrote custom laser-trapping data-acquisition \& analysis software. M.J.C conducted laser trapping control experiments, laser alignment, and data analysis. P.J.P. conceived of the laser trapping experiments as well as contributed to data analysis. P.B.R., B.E.S, X.Z., and P.J.P. wrote the manuscript, with input from all co-authors.%
%
\bibliographystyle{apsrev}
\bibliography{arXivCooling}

\section{\label{sec:App}Appendices}
\subsection{\label{sec:AppI}Appendix I: Cold Brownian Motion Temperature Extraction with Laser Trapping}
Power spectra from the quadrant photodiode (QPD) voltage traces were processed according to \citet{berg2004power-S1} and used to calibrate the QPD traces following the method of \citet{tolic2006calibration-29}. An experimental diffusion coefficient was then extracted by fitting the characteristic function for the experimental power spectra derived in \citet{berg2004power-S1}. Given that the temperature of the trapped particle is significantly different than the temperature sufficiently far from the laser focus, the particle-trap system is not isothermal and behaves according to nonequilibrium dynamics. Thus, equating the experimental diffusion coefficient to nonisothermal Brownian dynamics necessitates the application of cold Brownian motion (CBM), as derived by \citet{chakraborty2011generalised-7}. The CBM diffusion coefficient is then related to the CBM temperature by:
\begin{equation}
D_{CBM} = \frac{k_bT_{CBM}}{\gamma_{CBM}(T)},
\label{eqA1}
\end{equation}

\noindent where $D_{CBM}$ is the CBM diffusion coefficient, $k_b$ is Boltzmann's constant, $T_{CBM}$ is the CBM temperature, and $\gamma_{CBM}(T)$ is the CBM Stokes drag. To the leading order of the temperature increment or decrement $\Delta$T = (T$_p$ - T$_0$), the temperature-dependence of the viscosity on $T_{CBM}$ can be neglected, giving the effective temperature\cite{chakraborty2011generalised-7}
\begin{equation}
T_{CBM} = T_0 + \frac{5}{12}\Delta T.
\label{eqA2}
\end{equation}

For a temperature dependence of the solvent viscosity of the form $\eta(T) = \eta_\infty\exp\left[\frac{A}{(T-T_{VF})}\right]$, the CBM Stokes drag is given by
\begin{equation}
\gamma_{CBM}(T) = 6\pi R\eta_{CBM}(T),
\label{eqA3}
\end{equation}

\noindent where $R$ is the particle radius, and $\eta_{CBM}(T)$ is the temperature-dependent CBM viscosity that is related to the viscosity of the solvent at room temperature, $\eta_0$, by
\begin{widetext}
\begin{equation}
\frac{\eta_0}{\eta_{CBM}(T)} \approx 1 + \frac{193}{486}\left[\ln\left(\frac{\eta_0}{\eta_\infty}\right)\right]\left[\frac{\Delta T}{(T_0-T_{VF})}\right] - \left[\frac{56}{243}\ln\left(\frac{\eta_0}{\eta_\infty}\right) - \frac{12563}{118098}\ln^2\left(\frac{\eta_0}{\eta_\infty}\right)\right]\left[\frac{\Delta T}{(T_0-T_{VF})}\right]^2.
\label{eqA4}
\end{equation}
\end{widetext}

Equations \ref{eqA2}-\ref{eqA4} are then used in Eq. \ref{eqA1} to obtain $D_{CBM}$, which is subsequently compared to the experimental diffusion coefficient to determine the particle temperature T$_p$ (excluding the temperature discontinuity at the particle's surface from the Kapitza resistance\cite{cahill2003nanoscale-4}).

An alternative CBM temperature analysis using a semi-phenomenological expression for $D_{CBM}$ that approximately accounts for higher order terms in $\Delta$T (Eq. 15 of the supporting online materials of \citet{chakraborty2011generalised-7}) yields consistent results, indicating that these higher order corrections are negligible, for our purposes. For the experiments reported here, the VFT viscosity parameters were fit to experimental data and are as follows:

\begin{center}
\textbf{D$_2$O}\\
$\eta_\infty = 3.456*10^{-5}$ Pa$\cdot$s\\
$A = 478.6$ K\\
$T_{VF} = 160$ K,\\
\end{center}

and

\begin{center}
\textbf{DI water, PBS, DMEM}\\
$\eta_\infty = 2.664*10^{-5}$ Pa$\cdot$s\\
$A = 536.5$ K\\
$T_{VF} = 145.5$ K.\\
\end{center}

VFT viscosity parameters for DI water, PBS (0.01M, pH 7.4; Sigma P5368), and DMEM (1X, high glucose, pyruvate; Life Technologies Cat. \#11995-065) were assumed to be equivalent since it has been reported that water viscosity can be used for purposes of modeling particle transport in non-serum containing media\cite{hinderliter2010isdd-S5}.
\subsection{\label{sec:AppII}Appendix II: Laser Trap Temperature Extraction Error Analysis}
\begin{figure}[h]
	\begin{center}
		\includegraphics[width=8cm]{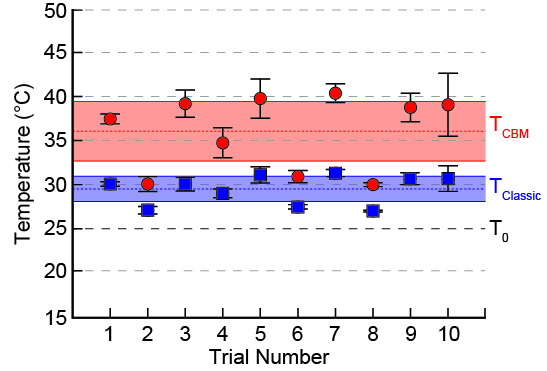}
	\end{center}
	\caption{\textbf{Laser trap temperature extraction error analysis.} A 1.01 $\mu$m diameter silica sphere trapped in D$_2$O using a trapping wavelength of 975 nm in a temperature controlled chamber held at T$_0$ = 25 $^\circ$C. QPD data was acquired 10 consecutive times at an irradiance of 5.9 MW/cm$^2$ and the temperature of the sphere was determined using (red circles, T$_{CBM}$) the cold/hot Brownian temperature extraction method detailed in Appendix \ref{sec:AppI}, and (blue squares, T$_{Classic}$) the classical isothermal assumption method. The error bars represent the standard deviation.}
	\label{fig:error}
\end{figure}
In order to determine the accuracy of the temperature extraction method presented here, silica beads (SS04N/9857, Bangs Laboratories) are used for their monodisperse size distribution (1010 nm diameter) and their known capacity to heat minimally when trapped with a NIR laser trap\cite{peterman2003laser-30}. The same experiment that was performed on YLF particles in this study was performed on the silica beads in D$_2$O using a trapping wavelength of 975 nm and an irradiance of 5.9 MW/cm$^2$. For this experiment, a temperature controlled chamber was used (RC-31, Warner Instruments), with the chamber held at T$_0$ = 25 $^\circ$C while the same bead was trapped and QPD data acquired 10 consecutive times. The cold Brownian analysis outlined above in Appendix \ref{sec:AppI} was then applied to the resulting QPD signals, and the temperature results are shown in red circles in Fig. \ref{fig:error}. The average calculated particle temperature (red dashed line) is T$_{CBM}$ = 35.9 $^\circ$C with a standard deviation (red shaded area) of $\pm$6.8 $^\circ$C. For comparison, the same data was also analyzed with the method assuming isothermal conditions, and the temperature results are shown in blue squares in Fig. \ref{fig:error}. The average calculated particle temperature (blue dashed line) is T$_{Classic}$ = 29.5 $^\circ$C with a standard deviation (blue shaded area) of $\pm$2.8 $^\circ$C.
\subsection{\label{sec:AppIII}Appendix III: Ratiometric Thermometry Calibration}
\begin{figure}[h]
	\begin{center}
		\includegraphics[width=7cm]{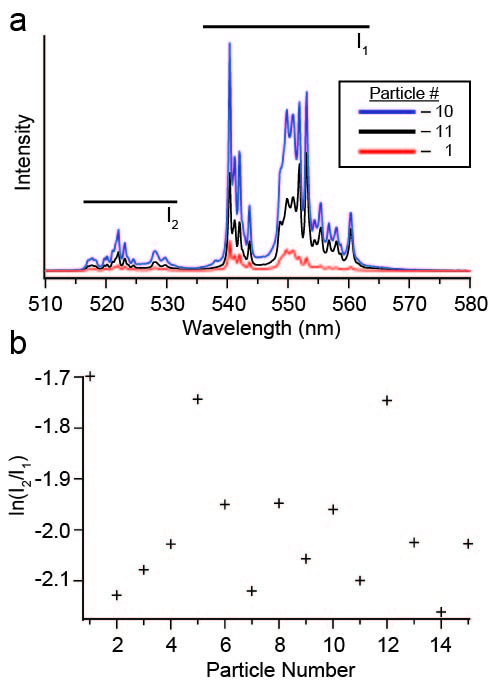}
	\end{center}
	\caption{\textbf{Ratiometric thermometry variability. a,} Select photoluminescence spectra from the measurements in (b) of multiple optically trapped, codoped 2\%Er$^{3+}$,10\%Yb$^{3+}$:YLF particles showing the variability in the integration regions I$_2$ and I$_1$, representing emission from Er$^{3+}$ energy states E$_2$ ($^2$H$_{11/2}$) and E$_1$ ($^4$S$_{3/2}$) to the ground state E$_{ground}$ ($^4$I$_{15/2}$), respectively. \textbf{b,} Natural logarithm of the ratio I$_2$/I$_1$ of different trapped particles showing the particle-to-particle variation.}
	\label{fig:vary}
\end{figure}
Calibration of the ratiometric thermometry on an individual particle is non-trivial in that there exist fluctuations in particle morphology and dopant concentration/distribution that affect the kinetics of the upconversion efficiency and overall spectral profile of the E$_1$ ($^4$S$_{3/2}$) and E$^2$ ($^2$H$_{11/2}$) to E$_{ground}$ ($^4$I$_{15/2}$) transitions (Fig. \ref{fig:vary}a). This means that for a given laser power, relative intensities will vary from particle to particle (Fig. \ref{fig:vary}b). It is, therefore, improbable that a comparison of the individual particles' emission ratio of the two transitions to a calibrated ensemble would yield an accurate temperature. For this reason, we use the Boltzmann distribution of these electronic states solely in a qualitative manner to confirm the trend of a decreasing or increasing temperature.
\begin{figure}[h]
	\begin{center}
		\includegraphics[width=8cm]{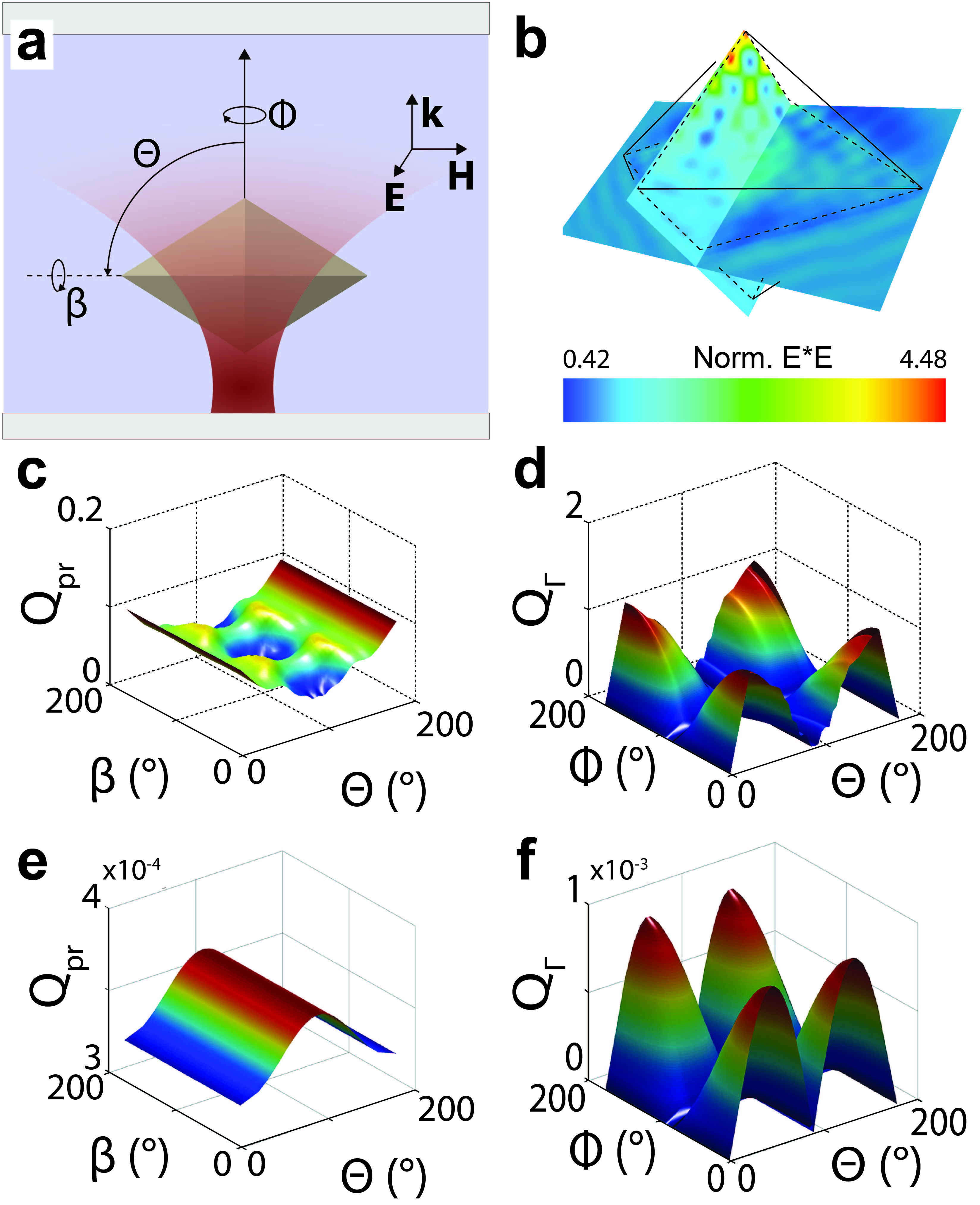}
	\end{center}
	\caption{\textbf{DDSCAT Simulations of Optically Trapped YLF. a,} Optically-trapped YLF particle in an aqueous fluid chamber. Angles are used to describe numerical discrete dipole approximation calculations in panels c-f. \textbf{b,} Cross sectional plot of the internal electric field magnitude within a YLF particle with dimensions 7.6 $\mu$m \& 3.9 $\mu$m at a trapping wavelength of 1020nm. \textbf{c,} Radiation pressure on a YLF particle shown in b. \textbf{d,} Torque calculations as a function of angular coordinates for the YLF particle shown in b. \textbf{e,} Radiation pressure on a 200 nm long YLF. \textbf{f,} Torque calculations as a function of angular coordinates for a 200 nm long YLF particle.}
	\label{fig:ddscat}
\end{figure}
\subsection{\label{sec:AppIV}Appendix IV: Yb$^{3+}$ Ion Spacing}
According to the XRD data presented in Fig. \ref{fig:character}c and detailed in Materials \& Methods, the lattice parameters of 10\%Yb$^{3+}$:YLF are a = 5.1641 {\AA} and c = 10.7177 {\AA} and the volume of the corresponding unit cell is V$_0$ = a$^2*$c = 2.85$*10^{-28}$ m$^3$. For 1 m$^3$ YLF, the number of unit cells is n = 1 m$^3$/V$_0$ = 3.5$*10^{27}$ m$^{-3}$. For each unit cell, there are 4 rare earth atoms. The total number of rare earth atoms in 1 m$^3$ is n$_t$ = 4n = 1.4$*10^{28}$  m$^{-3}$. For 10\%Yb$^{3+}$:YLF, the number of Yb atoms in 1 m$^3$ is n$_{Yb}$ = 0.1$*$n$_t$ = 1.4$*10^{27}$ m$^{-3}$. The linear density of Yb atoms is p = (n$_{Yb}$)$*(1/3)$ = 1.11$*10^8$ m$^{-1}$ and the average ion spacing is approximately l = 1/p = 8.9 {\AA}.
\subsection{\label{sec:AppV}Appendix V: Electromagnetic Simulations of Optically Trapped YLF}
Discrete Dipole Approximation software DDSCAT 7.3\cite{draine1994discrete-S7} was used to simulate laser trap interactions with ideal (non-truncated) YLF tetragonal bipyramids in water. For a bipyramid of length of 7.6 $\mu$m and base of 3.9 $\mu$m (Fig. \ref{fig:ddscat}b), a dipole density of 2267 dipoles/$\mu$m$^3$ (85,901 dipoles/octahedron) was used. At 1020 nm, refractive indices of YLF and water are 1.44831 and 1.327, respectively. Although YLF is a birefringent crystal, this has been ignored in the simulation based on the small birefringence ($\sim$0.02), which will only be reduced by the lattice mismatch of the many domains within a single octahedron. Also, the approximation of a laser trap by a plane wave becomes less appropriate as the size of the particle relative to the beam spot gets large. By varying $\Theta$, $\Phi$, and $\beta$ (see Fig. \ref{fig:ddscat}a) and calculating the radiation pressure and torque we are able to predict the optimal orientation for trapped particles since we assume the particle will orient so as to minimize these forces and maximize the trapping potential. One intriguing result of the simulations is the particle size dependence of orientation. For example, a tetragonal bipyramid with a length of 7.6 $\mu$m is expected to align with the long axis perpendicular to both k and E, or $\Theta$=90$^\circ$, $\Phi$=90$^\circ$, and $\beta$=0$^\circ$ (Fig. \ref{fig:ddscat}c/d), whereas a similar particle with a length of 200 nm is predicted to align its long axis to be parallel with the direction of propagation (Fig. \ref{fig:ddscat}e/f). It is also notable that the particle is able generate significant internal and near fields as demonstrated in Fig. \ref{fig:ddscat}b.
\end{document}